\documentclass[apl,reprint]{revtex4-1}
\usepackage{graphicx}
\usepackage{bm,braket,color,ulem}
\usepackage{amssymb}

\begin{document}

\title{Near-field microwave imaging of inhomogeneous K$_x$Fe$_y$Se$_2$\\: separation of topographic and electric features}

\author{Hideyuki Takahashi$^{1}$, Yoshinori Imai$^{2}$, Atsutaka Maeda$^{2}$}
\affiliation{
$^{1}$Organization of Advanced Science and Technology, Kobe University, 1-1, Rokkodai, Nada, Kobe 657-8501, Japan\\
$^{2}$Department of Basic Science, the University of Tokyo, 3-8-1 Komaba, Meguro-ku, Tokyo 153-8902, Japan}

\date{\today}

\begin{abstract}
It is important for modern scanning microwave microscopes to overcome the effect of the surface roughness. Here, we report microwave conductivity imaging of the phase-separated iron chalcogenide  K$_x$Fe$_y$Se$_2$ ($x=0.8$, $y=1.6$-$2$), in which electric conductivity-induced contrast is distinguished from topography-induced contrast using a combination of a scanning tunneling microscope and a scanning microwave microscope (STM-SMM). We observed the characteristic modulation of the local electric property that originates from the mesoscopic phase separation of the metallic and semiconducting phases in two different scanning modes: constant current (CC) mode and constant $Q$ (CQ) mode. In particular, CQ scanning is useful because we obtain a qualitative image in which the topographic contrast is largely eliminated without degradation of the spatial resolution.
\end{abstract}

\maketitle
The study of the intrinsic inhomogeneity of materials has been crucial in modern condensed matter physics. For example, in materials such as high-$T_{\mathrm{c}}$ cuprate and Fe-based superconductors, clarifying the relation between microscopic (or mesoscopic) inhomogeneity and macroscopic material properties has been a serious issue \cite{Pan2001,Allan2013,Liu2012,Ding2013}. For this purpose, scanning microwave microscopes (SMMs), which utilizes a near-field microwave electromagnetic field to detect local microwave responses can be useful tools \cite{Wei1996}. This tool enables us to study the local electrodynamics of elementary excitation via the local electric property at microwave frequencies.

Recently, the SMM topographic spatial resolution has been significantly improved by combining this tool with a scanning tunneling microscope (STM) \cite{Imtiaz2003,Machida2009,Lee2010} or an atomic force microscope (AFM) \cite{Lai2007}. 
An STM-SMM can obtain an atomic resolution image of highly ordered pyrolytic graphite (HOPG) by mapping  the energy loss caused by the tunneling junction impedance \cite{Lee2010,Reznik2014}.
However, to enhance the value of the modern SMM as a convenient tool for materials research, its capability for complex conductivity mapping should be improved.
And necessarily, we must address one of the most serious problems: separation of the topographic and electric information.
Unwanted topographic information inevitably appears in microwave images for both STM-SMM and AFM-SMM \cite{Imtiaz2006,Imtiaz2007}. 
Topography-induced contrast often becomes a large background and obscures the electric inhomogeneity.
Therefore, it is important to develop a technique which enable us to remove topographic contrast from microwave images to obtain intrinsic information about the electric inhomogeneity.

In this letter, we report near-field microwave imaging of phase-separated iron chalcogenide K$_x$Fe$_y$Se$_2$ ($x=0.8$, $y=1.6$-$2$) \cite{Guo2010}, in which the electric and topographic contrasts are distinguished using an STM-SMM.
Below the structural transition temperature of approximately $580\ \mathrm{K}$ this compound separates into a major K$_2$Fe$_4$Se$_5$ phase and an Fe-rich phase because of the disorder-to-order transition of the Fe vacancy sites \cite{Bao2011}.
Scanning electron microscopy/energy-dispersive X-ray spectroscopy (SEM/EDS) has revealed that the Fe-rich phase forms an orthogonally alligned web-like network and separates the majority phase into mesoscopic domains \cite{Liu2012}. 
Accumulated transport studies have indirectly suggested that the Fe-rich minority phase is metallic ($\sigma\sim 10^5\ \mathrm{S/m}$, this phase becomes superconducting below $31\ \mathrm{K}$) and that the majority phase is semiconducting ($\sigma\sim 10^2\ \mathrm{S/m}$) \cite{Shen2011,Ding2013}.
These features make this compounds suitable for demonstrating the ability of SMM and studying the effect of surface roughness on microwave images.

The K$_x$Fe$_y$Se$_2$ bulk crystal was grown using the self-flux method \cite{Guo2010}. 
First, an FeSe precursor was prepared by melting Fe (99.999\%) and Se (99.9999\%) grains at 1050$^\circ$C. 
Then, K and FeSe were placed into an alumina crucible with a mixing ratio of K:FeSe=1:2. 
The crucible was placed in a stainless steel tube that was sealed with a cap union nut. These processes were performed in a glove box filled with high-purity argon gas. 
The mixture was heated to 1030$^\circ$C and maintained at that temperature for 2 h. 
Afterwards, the melt was slowly cooled down to 400$^\circ$C over 40 h before the furnace was shut down. 
Then, we sealed the obtained as-grown crystals into a quartz tube under vacuum, which was annealed at 400$^\circ$C for 1 h, followed by quenching in cold water \cite{Liu2012,Ding2013}. 
The sample was cleaved and set on the sample stage of the SMM head under a dry N$_2$ atmosphere, because this compound is moderately sensitive to moisture. 
Then, the SMM chamber was evacuated and filled with dry He gas.

\begin{figure}[tb]
	\begin{center}
		\includegraphics[width=0.9\hsize]{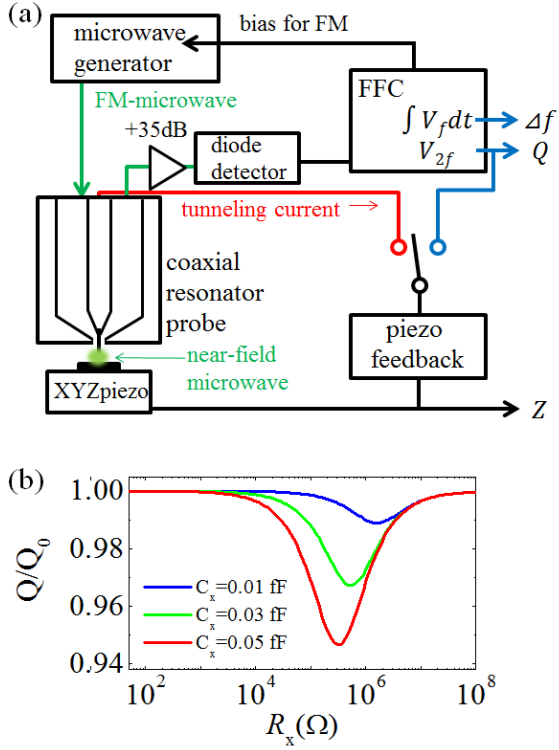}
	\end{center}
\caption{(a) Schematic of the SMM measurement system. The feedback signal for tip-sample distance control is selected from the tunneling current or $V_{2f}$. (b) Calculated $R_x$ and $C_x$ dependence of $Q$.}
\label{fig1}
\end{figure}

Figure \ref{fig1}(a) presents a schematic illustration of our coaxial resonator probe and measurement system. The resonant frequency of the probe is $f_0=\omega/2\pi=10.7 \ \mathrm{GHz}$, and the resonant mode is the transverse electromagnetic (TEM) $3\lambda/4$ mode, in contrast with most previous studies, which used the TEM $\lambda/4$ mode \cite{Wei1996,Lee2010}. A metal tip is connected to the center conductor and protrudes out of the aperture of the outer conductor to detect the tunneling current and near-field microwave response. We used electrochemically etched Pt-Ir alloy wires whose curvature radius at the end, $r_{\mathrm{tip}}$, was typically 100 nm. The tunneling current is amplified and sent to an ordinary STM feedback circuit to control the distance between the tip and sample, $h$. By keeping the tunneling current constant during a scan, the surface topography and distribution of the local microwave response can be obtained simultaneously. The frequency-modulated (FM) microwave signal transmitted through the resonator probe is converted into a voltage signal at the diode detector and then processed at the frequency-feedback circuit (FFC) \cite{Imtiaz2006, Machida2009}. The diode dectector output mainly contains a signal at the modulating frequency, $V_f$ and its harmonics $V_{2f}$. The FFC keeps the microwave signal source locked to a resonant frequency using an integral control that uses $V_f$ as a feedback signal. We can evaluate the quality factor, $Q$, and frequency shift $\Delta f$ of the probe from $V_{2f}$ and $\int V_f dt$, respectively. The unloaded $Q$ is $Q_0=1200$-$1300$, which is more than twice as high as previously reported values \cite{Imtiaz2003, Lee2010}. More details about the instrumentation will be published elsewhere \cite{HTunpublished}

It is helpful to use the lumped element circuit model to understand the behavior of the resonator probe. 
Our TEM $3\lambda/4$ coaxial resonator is equivalent to an RLC parallel resonant circuit with $R\sim40\ \mathrm{k\Omega}$ and $C\sim0.5\ \mathrm{pF}$. 
As the tip approaches the sample, the load impedance changes from an infinite value to $R_x +1/j\omega C_x$, where $R_x$ is the near-field impedance of the sample and $C_x$ is the coupling capacitance between the tip and sample.
For conducting samples, $R_x$ is on the order of $R_x=\rho/r_{\mathrm{tip}}$, where $\rho$ is the dc resistivity. $C_x$ is calculated to be on the order of 0.01 fF by applying a parallel-plate approximation ($C_x \approx \epsilon_0 \pi r_{\mathrm{tip}}^2/h$, where $\epsilon_0$ is the dielectric permeability in vacuum). 
Figure 1(b) shows the calculated $Q$ of the equivalent circuit as a function of the load impedance. 
$Q$ exhibits a minimum value when $\omega C_x R_x =1$ and exhibits a monotonous $C_x$ dependence.
These behaviors are qualitatively consistent with previous experiments and finite-element simulations \cite{Hyun2002,Okazaki2007}.

\begin{figure}[tb]
	\begin{center}
		\includegraphics[width=0.9\hsize]{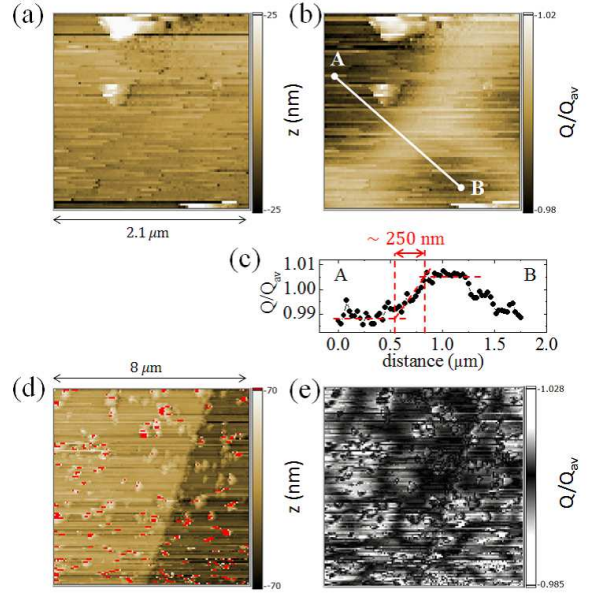}
	\end{center}
\caption{(a) STM topograph of K$_x$Fe$_y$Se$_2$. (b) $Q$ image of the same area as (a). $Q_{\mathrm{av}}$ denotes the average value of $Q$ over all the pixels. (c) A linecut from point A to B in (b). (d) STM topograph of the large region. The height of the step in the topograph is 40 nm. (e) $Q$ image of the same area as (d). In this image, the region where $Q/Q_{\mathrm{av}}\sim1.009$ is emphasized in black for clarity.}
\end{figure}
Figures 2(a) and 2(b) show the surface topograph and spatial dependence of $Q$ of the probe simultaneously acquired in the CC mode at room temperature. 
We can observe two types of changes in the $Q$ image. 
One type of change occurs in areas where a topographic change is also observed, and the other type occurs in smooth areas. 
A change in $Q$ following a topographic change is frequently observed in SMM \cite{Lai2007,Imtiaz2006} and is attributed to the abrupt change of $C_x$, which is very sensitive to $h$ and the geometry of the tip and sample.
When the tip goes up on a grain-like structure, $C_x$ decreases because the effective distance between the tip and sample increases.
In this situation, $Q$ becomes large as observed in Fig. 1(b).
However, changes in smooth regions should be related to the inhomogeneous electric property of K$_x$Fe$_y$Se$_2$, i.e., the mesoscopic phase separation of the metallic and semiconducting phase. 
Using $\rho$ of each phase, $R_x$s of the metallic and the semiconducting phase are estimated to be $\sim 10^2\ \mathrm{\Omega}$ and $\sim 10^5\ \mathrm{\Omega}$, respectively. 
These values are less than $R_x$, where $\omega C_x R_x =1$, as shown in Fig. 1(b).
In this regime, $Q$ monotonously increases as $R_x$ decreases.
Therefore, we concluded that the high-$Q$ region in Fig. 2(b) corresponds to the metallic phase. 

Figures 2(d) and 2(e) present the topograph and $Q$ image of the larger area. 
There is a lot of debris, which may have resulted from the chemical reaction between K$^+$ ions near the surface and contaminated moisture during the sample mounting process; these debris disturb the stable CC scanning and cause many streaks in the images. 
However, we can observe contrast in the $Q$ image, which is not related to a topographic change.
In Fig. 2(e) the metallic phase corresponds to the black region and forms the web-like network that spreads over the entire sample surface. Consequently, the semiconducting phase is separated into island-like domains with typical dimensions of $2\ \mathrm{\mu m}\times 2\ \mathrm{\mu m}$. 
These features of the metallic and semiconducting phases agree well with those of the minority and major K$_2$Fe$_4$Se$_5$ phase observed by SEM/EDS, respectively \cite{Liu2012,Ding2013}. 
This finding indicates that we have obtained direct evidence that the minority phase-forming microstructure is indeed metallic, as inferred from the results of indirect transport studies.

Next, we present the images acquired using another scanning mode. 
Generally, in scanning probe microscopy (SPM), precise control of tip-sample distance is achieved by keeping the signals that are observed only in the vicinity of the sample surface and change rapidly as the probe gets away from it constant.
For SMM, we can use $Q$ and $\Delta f$ as feedback signals instead of the tunneling current because these quantities also change monotonously when $h$ changes.
If the sample is electrically homogeneous, one can keep $h$ almost constant by keeping $Q$ or $\Delta f$ constant.
It has been demonstrated that topographic imaging of a conducting sample is possible in the constant-frequency mode \cite{AKim2003}.
This scanning modes can also be used to observe electrically inhomogeneous samples.
In that case, because $Q$ or $\Delta f$ depends on the local electric property of the sample, it is expected that we can obtain information about the inhomogeneous electric property using a trace of the tip movement when the sample has a sufficiently smooth surface.

In our setup, we cannot avoid making the phase-locked loop inside the microwave generator open to control the frequency modulation using an external circuit, which
results in a drift of the center frequency of the FM-microwave, consequently, we cannot operate our SMM in the constant-frequency mode.
We thus selected $Q$ as the feedback signal.
The tunneling current channel is isolated from the feedback circuit while the SMM is operated in the CQ mode.
However, the tunneling current is kept monitored so that the tip does not contact the sample because the change in $Q$ near the surface is not as steep as that of the tunneling current.

\begin{figure}[tb]
	\begin{center}
		\includegraphics[width=0.9\hsize]{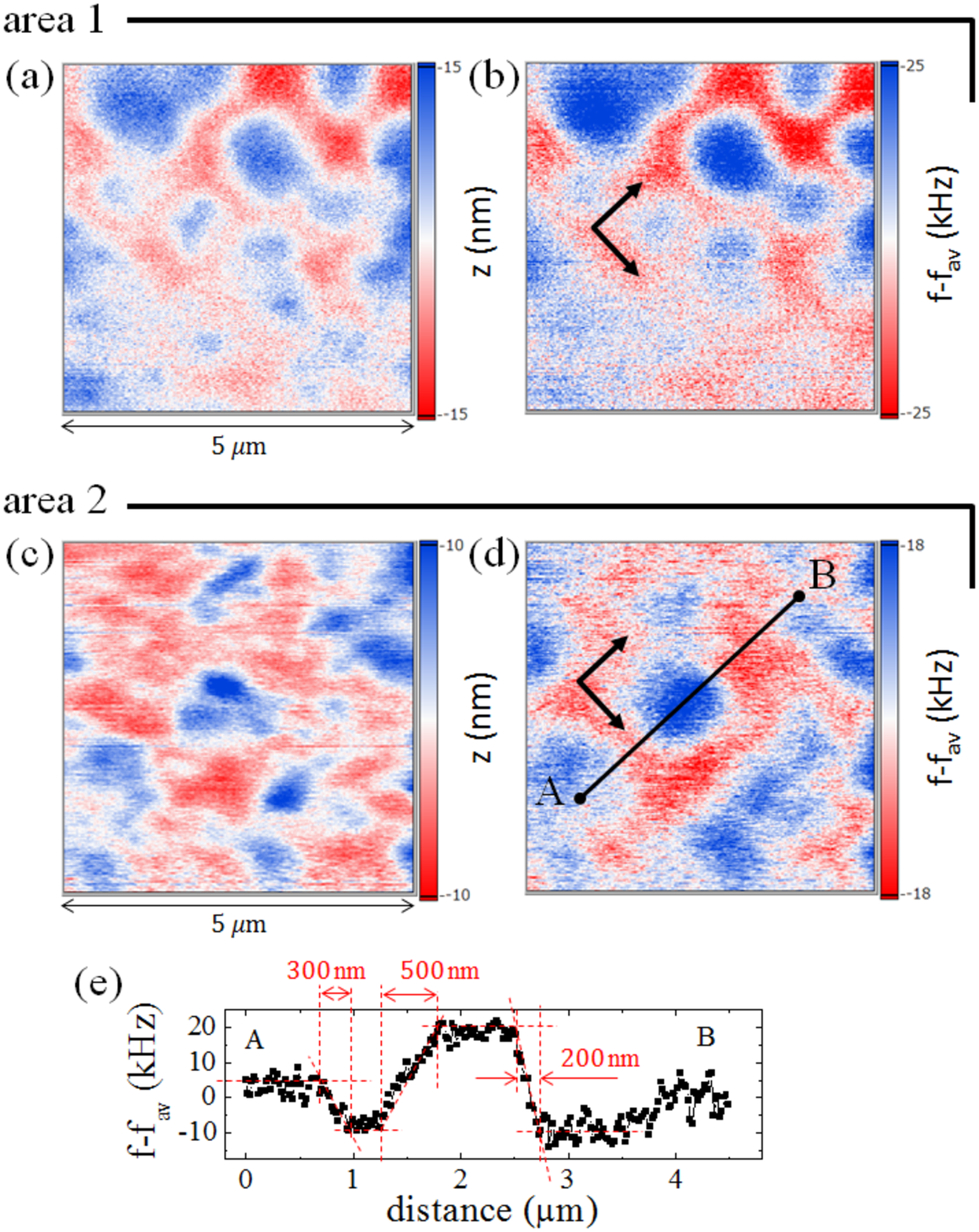}
	\end{center}
\caption{(a)-(d) $z$ and $\Delta f$ images acquired in the CQ mode in two different areas. In the $\Delta f$ images, the effect of the frequency drift of the microwave generator is corrected by subtracting the average value in each line. (e) A linecut from the point A to B in the $\Delta f$ image in (d). The black arrows indicate the direction in which the low-frequency region spreads.}
\label{fig3}
\end{figure}

Figures \ref{fig3}(a)-(d) show the $z$ (trace of tip movement) and $\Delta f$ images of two different areas acquired in the CQ mode.
The set point of $Q$, $Q_{\mathrm{set}}$, was determined such that $h$ was maintained at approximately $20\ \mathrm{nm}$ on average.
The same scanning direction was used for two regions.
In Fig. 3(a), which was acquired from area 1, we can see domain structures in the $z$ image.
The blue domain, in which $z$ is large, corresponds to a lossy semiconducting region when we interpret the origin of the contrast as electric inhomogeneity.
In contrast, the z image in area 2, in which the domain structure is blurred, is very different (fig. 3(c)).
However, it is surprising that the two $\Delta f$ images  (Fig. 3(b) and Fig. 3(d)) have common features.
The domain structures are clearly defined and the low-frequency regions are spread in the same orthogonal directions, as indicated by the black arrows; this behavior is characteristic of the phase separation in K$_x$Fe$_y$Se$_2$.

The reason why the $\Delta f$ image reflects electric inhomogeneity more clearly than the $z$ image is as follows.
Here, we refer to the tip height at which $Q(h)=Q_{\mathrm{set}}$ in the  I phase as $h_i$. 
In the CQ mode, the tip moves vertically detecting the surface geometry along with electric inhomogeneity (see Fig. 4).
Thus the plotted quantity in the $z$ image is the superposition of the surface topograph and tip height, $z(r)+h_i$.
The difference between Fig. 3(a) and Fig. 3(c) is attributed to the surface topograph.
However, $\Delta f$ mainly depends on $h_i$ and exhibits constant values in each separated phase.
$\Delta f$ exhibits significant changes only when the tip crosses the boundary of regions with different conductivities.
One might be worried that the fluctuation in $C_x$ during scanning, which is caused by the geometry of the tip and sample, affects the $\Delta f$ image.
However, $C_x$ is at least one order of magnitude less than that obtained using tunneling current feedback because we set $Q_{\mathrm{set}}$ such that $h$ is approximately 20 nm.
Thus, the effect of the fluctuation of $C_x$ is negligibly small when the surface geometry does not abruptly change in a scale shorter than $r_{\mathrm{tip}}$, and we can obtain qualitative images in which the topographic information is largely eliminated.
\begin{figure}[tb]
	\begin{center}
		\includegraphics[width=0.9\hsize]{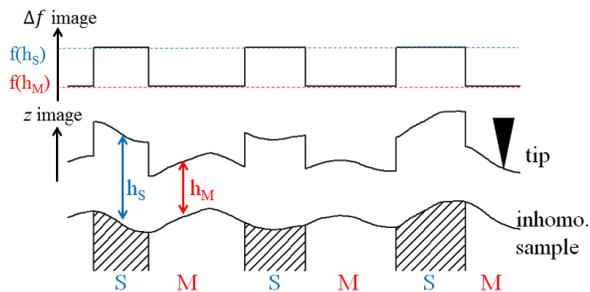}
	\end{center}
\caption{Schematic illustration of CQ scanning mode. S and M denote the semiconducting and metallic phase, respectively.}
\label{fig4}
\end{figure}

The CQ mode has great advantages in visualizing electric inhomogeneity.
Thus far, the preparation of a very flat surface has been required for SMM measurements to avoid the contrast induced by the surface roughness \cite{Imtiaz2007}.
However, our results demonstrate that one can determine the electric inhomogeneity using the CQ scanning mode even for samples that exhibit some roughness.
In addition, we do not necessarily need the functionality of STM for this mode.
Thus, we can operate the SMM even when it is difficult to stably detect the tunneling current because of improper surface condition.
In fact, although we could not operate the SMM in the CC mode for the sample used in Fig. 3, we obtained the images of the intrinsic electric inhomogeneity in the CQ mode.
Since this method is qualitative, we should be careful in the application for more general case with multiple phases. Particularly, when the sample contains both the metallic and the insulating region across the minimum in Fig. 1(b), we need additional quantitative information to interpret the obtained images. In such a situation, it is important to measure the $h$ dependences of $Q$ and $\Delta f$ in each phase in advance.

Because we could separate the topographic and electric contrast, we could evaluate the electric spatial resolution of our SMM, which is defined as the ability to distinguish two points that have different conductivities as a contrast in the microwave image.
Figures 2(c) and 3(e) show the linecuts of microwave images shown in Fig. 2(b) and Fig. 3(d), respectively.
As the tip crossed the boundary between different phases, the change in the microwave response occured within a width of $\sim$200 nm at the narrowest part.
This result indicates that the spatial resolution is no worse than 200 nm.
It is noteworthy that the spatial resolution did not degrade even in the CQ mode, in which the tip height is higher than that in the CC mode.
In the infinitesimal vertical dipole model, the characteristic length scale of localization of the electromagnetic field at the tip end is approximately $r_{\mathrm{tip}}+h$ \cite{Imtiaz2006}.
Therefore, if $h\ll r_{\mathrm{tip}}$, as in the current situation, SMM maintains a high spatial resolution.
For further improvement of the spatial resolution, sensitivity improvement of the resonator probe is required along with the use of sharper tips because the coupling between the tip and sample becomes weaker as $r_{\mathrm{tip}}$ decreases. 

To summarize, we reported near-field microwave imaging of phase-separated K$_x$Fe$_y$Se$_2$ using STM-SMM.
We observed web-like metallic network and island-like semiconducting domains that originate from the mesoscopic phase separation in two different scanning modes.
In the CC mode, we needed to carefully examine whether the origin of the contrast in the microwave image was topographic or electric.
However, in the CQ mode, we obtained a $\Delta f$ image in which the topographic contrast was largely eliminated without degradation of the electric spatial resolution.
\newline
\newline
\indent This work has been supported by a Grant-in-Aid for Scientific Research(A) (Grants. No. 23244070) from the Ministry of Education, Culture, Sports, Science, and Technology of Japan. H. Takahashi also thanks the Japan Society for the Promotion of Science for financial support.

\end{document}